\documentclass[11pt,a4paper,fleqn]{article}
\usepackage{epsfig}
\usepackage{amsmath}
\usepackage{amssymb}
\usepackage{amsfonts}
\usepackage{pifont}
\begin{document}
\title{Cosmic Rays above $10^{14}$ eV}
\author{Murat BORATAV\\LPNHE - Universit\'e Paris 6\\4 Place Jussieu, 75005 Paris (France)\and Alan A. WATSON\\Department of Physics and Astronomy\\University of Leeds, Leeds LS2 9JT (United Kingdom)}
\maketitle
\begin{abstract}
We briefly review the status of cosmic ray studies between $10^{14}$~eV and the highest observed energies, namely a few times $10^{20}$~eV. Because of the rather low incident fluxes in this energy range, the studies mostly rely on ground based, large aperture detectors reconstructing the cosmic ray's properties through the detection of the air-showers they generate by interacting with the atmosphere. We stress the fact that many issues such as the chemical composition of the cosmic rays, their acceleration mechanisms, the structures displayed in their energy spectrum are mostly open questions which may be answered by the next generation of experiments.
\end{abstract}

\section{Introduction}

The earth is continuously bombarded by a stream of high energy particles which arrive uniformly from all directions.  These particles are cosmic rays and were discovered in 1912 by Hess through a series of balloon flights in which he carried electrometers to over 5000~m.  We now know that the cosmic ray spectrum extends from 1 GeV to above $10^{20}$ eV (or 100 EeV) but we have only a rudimentary understanding of where cosmic rays come from.  The difficulty is that the particles are stripped nuclei of atoms: consequently the galactic magnetic fields, typically a few $\mu$G, are sufficiently strong to scramble their paths except at the very highest energies.  In this short review we will describe the properties of cosmic rays from $10^{14}$~eV up to the highest energy so far found, $3\times 10^{20}$~eV.  Below $10^{14}$~eV the flux of particles is sufficiently high that individual nuclei can be studied by flying complex detector packages in balloons.  It is found that the majority of particles are the nuclei of the common elements and that around 1 GeV the frequency distribution is strikingly similar to that found in ordinary material in the solar system.  The striking exception is the abundance of elements such as Li, Be and B which are over-abundant because of the fragmentation of heavier nuclei against inter-stellar hydrogen.

Above $10^{14}$~eV the techniques used to study the particles employ a phenomenon discovered by Auger and his group in 1938 \cite{Auger}.  When a high energy particle enters the atmosphere it initiates a cascade or air shower which is large enough and sufficiently penetrating to reach ground level.  At $10^{15}$~eV about $10^6$ particles, mainly electrons, are spread out over about a hectare.  With a small number of judiciously positioned detectors it is possible to measure the direction of the incoming shower to within about 1 degree using the relative arrival times of the shower at the detectors.  The size of an air shower array depends on the energy region which is being studied.  To explore the region from $10^{14}$ to $10^{16}$~eV an area of $4\times 10^4$~m$^2$ is typical but at the highest energies, where the rate is measured in particles per km$^2$ per century, much large areas must be instrumented.  The Pierre Auger Observatory, which will collect an unprecedented number of events above $10^{19}$~eV, will cover an area 30 times the size of the city of Paris.  

\section{Cosmic Rays in the Region of the Knee}

The cosmic ray energy spectrum is nearly featureless lacking the lines or dips which would characterise an electromagnetic spectrum covering so many decades.  It is often described in terms of a power law which fits the data over many decades.  Thus the differential spectrum is of the form $dI=kE^{-\gamma}dE$. One of the most prominent features is  the steepening of the slope of the energy spectrum from around $\gamma = 2.7$ to $\gamma  = 3.1$ at an energy of $3\times 10^{15}$~eV.  This is known as the `knee' of the spectrum.  At an energy above $10^{18}$~eV the spectrum flattens at what is called the `ankle'.  The knee in the spectrum was first deduced from observations of the shower size spectrum made in 1956 but it remains unclear as to what is the cause of this spectral steepening.

It is currently believed that cosmic rays are accelerated in a process called diffusive shock acceleration.  Suitable astrophysical shocks occur in supernovae explosions and particles of the interstellar medium gain energy as they are repeatedly overtaken by the expanding shock wave.  Most work (e.g. \cite{Ellison}) leads to the conclusion that the accelerated particles will have a spectrum close to $E^{-2.1}$,  considerably flatter than the measured value.  The maximum energy reached in the shocks associated with supernova remnants is close to $Z\times 10^{14}$~eV (see also \cite{Lagage}) and at the maximum energy the spectrum of a species falls abruptly.  This model can be reconciled, to some extent, with observations when the production spectrum is folded with the energy-dependence of the lifetime of cosmic rays in the galaxy.  Analysis of the abundance of fragmentation nuclei with energy suggests that the lifetime varies as $E^{-0.6}$. Thus the observed and predicted spectral slopes ($\gamma= 2.7$) can be reconciled.  However this picture cannot hold to energies very far beyond the knee as the lifetime of the particles in the galaxy is predicted to be so short that strong anisotropies should be seen above $10^{16}$~eV whereas the observed uniformity is within 1\%.  This difficulty has led Swordy \cite{Swordy} to suggest that there is a minimum propagation path length of 0.013~g~cm$^{-2}$ or about 3~kpc.  Alternatively Hillas \cite{Hillas} has suggested that the trapping time varies as $E^{-0.33}$ (as found with Kolmogorov turbulence) and that the energy spectrum of particles from the supernovae is much steeper than $E^{-2.1}$ but extends to higher energies.

It would help greatly to constrain these and other theories if the energy spectrum was measured more exactly and, in particular, if the way in which the mass composition changes in the decades before and after the knee was better known.  Significant progress has been made on these problems in recent years, but they are far from solved. The shower observables at ground level (sometimes mountain altitudes when observing in the knee region) are the number of muons, the number of electrons, the Cherenkov light signal produced as the electrons (mainly) propagate through the atmosphere and, but measured at only a small number of installations, the number and energy spectrum of hadrons.  To go from ground parameters to the primary energy requires recourse to a Monte Carlo calculation in which the particle interactions are followed stochastically, with assumptions being made about the particle physics characteristics of the interactions.  At $10^{15}$~eV one is well above accelerator energies and therefore important variables such as cross-sections, inelasticity and multiplicity must be inferred from extrapolation, often guided by a model.  Nucleon-nucleus, nucleus-nucleus and pion-nucleus collisions are all important.  Until their energy falls below about $10^{12}$~eV, the charged pions are more likely to interact and further feed the hadronic part of the cascade.  The neutral pions, which decay almost instantaneously, fuel the electromagnetic cascade which can be calculated very exactly.

\begin{figure}[!htb]
\begin{center}  
\epsfig{file=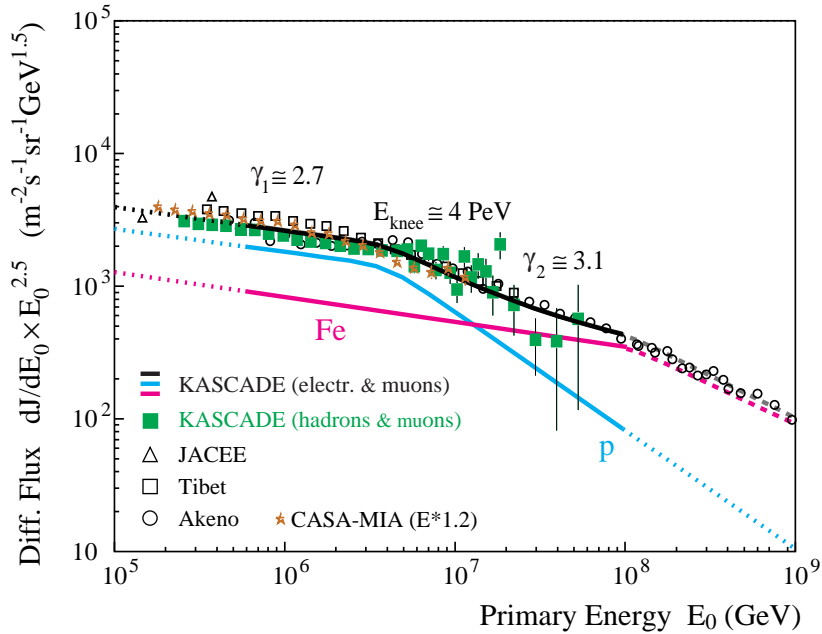,width=11cm}
\end{center}  
\caption{\small Cosmic ray energy spectrum from the KASCADE experiment \cite{Haungs}.\label{kascade}}
\end{figure}
Extracting the energy is not straightforward and a model must be adopted to relate the observed numbers of electrons and muons to the primary energy.  Currently the model which appears to fit much of the data in the knee region is the QGSJET model \cite{QGSJet} based on the quark-gluon-string model of particle interactions.  This model is required to extrapolate accelerator observations, which are largely made in the central rapidity region, to the forward region where observations are lacking but which are of most importance to air shower simulations.  A number of detailed studies \cite{Erlykin} find that the QGSJET model can describe much of the data on muons and electrons.  However for the hadronic component, measured above 50~GeV with high accuracy in the KASCADE experiment, it is found that the model predicts more hadrons than are observed in showers produced by primaries of  $10^{16}$~eV.  Only if the primaries at this energy were heavier than Fe could the data be reconciled with the QGSJET model.  Such a solution is regarded as unphysical because of the decline in the natural abundance of elements heavier than Fe.  A convincing representation of the energy spectrum from the KASCADE data \cite{Haungs} is shown in figure \ref{kascade}.  This spectrum agrees with data from other recent experiments such as CASA-MIA \cite{Glasmacher}, HEGRA \cite{Rohring} or CASA-BLANCA \cite{Fortson} when the same model is used to interpret the ground level data.  Using other models can change the intensity at a given energy by as much as 30\% (see \cite{Watson} for a discussion).  Generally different models do not alter the slope before or after the knee which is close to $3\times 10^{15}$~eV by most estimates.
\begin{figure}[!htb]
\begin{center}  
\epsfig{file=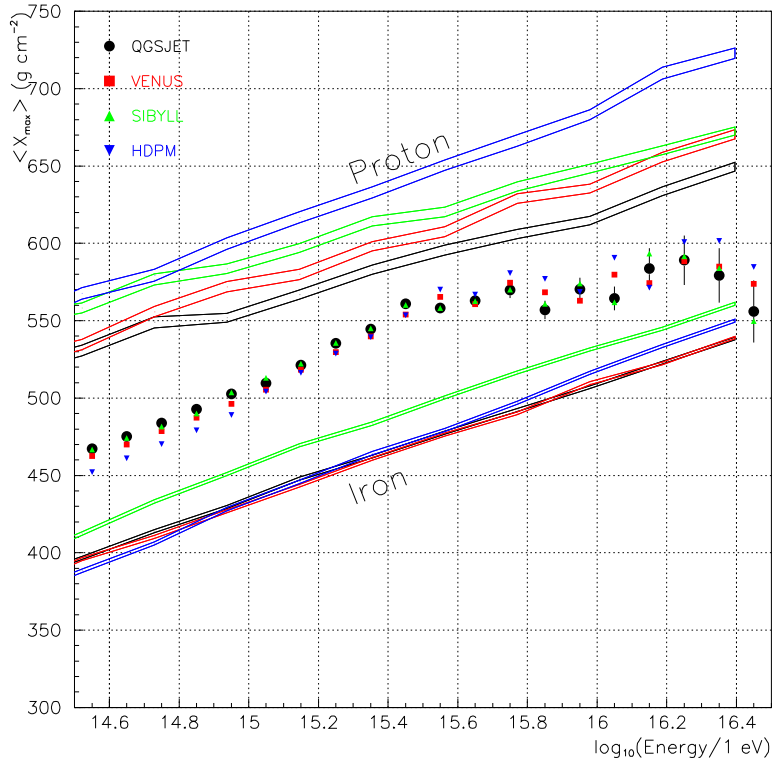,width=11cm}
\end{center}  
\caption{\small Variation of the depth of shower maximum as a function of energy.\label{fowler2}}
\end{figure}

\begin{figure}[!htb]
\begin{center}  
\epsfig{file=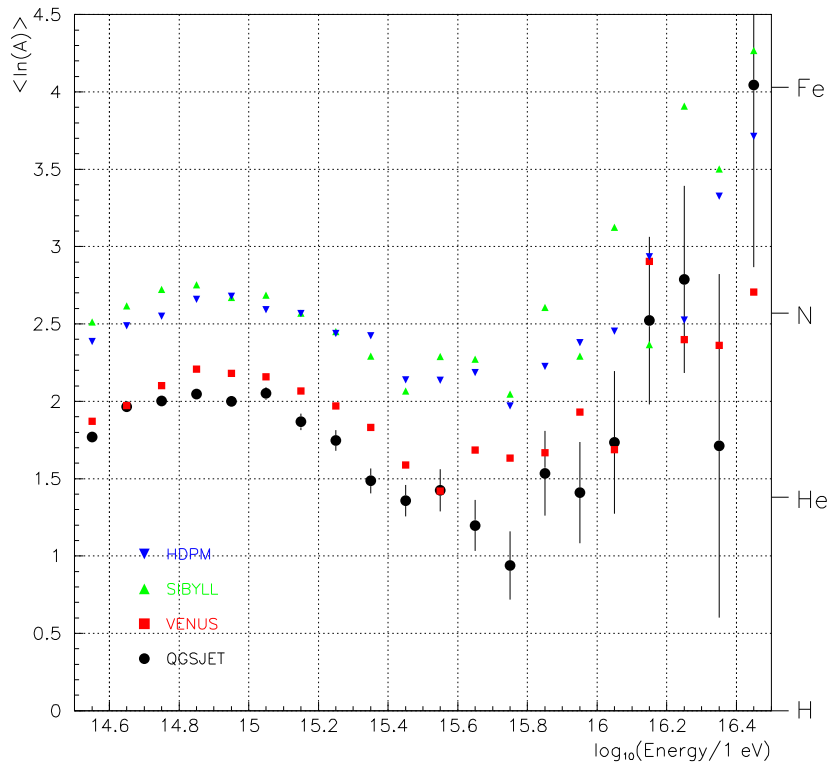,width=11cm}
\end{center}  
\caption{\small Variation of the mean mass as a function of energy \cite{Fortson}.\label{fowler3}}
\end{figure}

There are many approaches to determining the primary mass in the region of the knee \cite{Watson}. Here there is space only to mention one of these methods which has been developed recently and which looks rather promising. Many of the electrons in a air-shower have a speed high enough for them to radiate part of their energy as Cherenkov light and this can be detected above the night sky background on clear moonless nights.  Model calculations have shown \cite{Patterson} that the Cherenkov light signal at 100~m (and further) from the axis is a good measure of primary energy while the slope of the Cherenkov lateral distribution is sensitive to the depth of shower maximum, $X_{\text{max}}$.   Although the conversion from the observables to the energy and the depth of maximum is mass dependent, methods have been developed to circumvent this difficulty to some extent.  Cherenkov light emission from showers has been studied at a number of experiments \cite{Dickinson,Rohring2,Fortson} but, for illustrative purposes, only data from CASA-BLANCA will be described. This experiment used the CASA array of scintillators \cite{Borione} to find the shower cores and an array of 144 photomultipliers to measure the Cherenkov light parameters.  Details of the analysis have been given in a series of articles \cite{Fortson} and results are shown in figure \ref{fowler2}.  What is impressive is that for a range of four models (HPDM, QGSJET, SIBYLL and VENUS) and a mass mixture of protons, nitrogen and iron, the derived values of $X_{\text{max}}$  are quite model independent.  At $10^{15}$~eV where $X_{\text{max}}$ is about 500~g~cm$^{-2}$, the spread in the inferred value is only 10~g~cm$^{-2}$.  However for a particular mass the values predicted for $X_{\text{max}}$ show considerable spread.  Again taking $10^{15}$~eV as a reference, the difference in $X_{\text{max}}$ predicted by different models for protons is 50~g~cm$^{-2}$ and for iron about 30~g~cm$^{-2}$.  This implies that the extracted mean mass is model dependent, as is seen in figure \ref{fowler3} where the results of Fortson \emph{et al} \cite{Fortson}, deduced from their measurements of $X_{\text{max}}$ as a function of energy, are shown.  Here $<\ln A>$ is plotted against energy for various models.  Mean $\ln A$ is an appropriate variable as at a given energy $X_{\text{max}}$  is known to vary with $\ln A$ (Linsley 1977).

The trends in the data are identical for the different shower models.  The mean logarithmic mass becomes lighter in the knee region than at lower energies before becoming heavier again at higher energies.  Unfortunately the data at balloon energies are not sufficiently accurate to guide the choice of models.  At $10^{14}$~eV/particle the measured value of $<\ln A>$ is $1.8\pm 0.4$ and at $10^{15}$~eV (based on only 4 directly measured events) the corresponding value is $2\pm 1$.  However the KASCADE analysis of electrons, muons and hadrons is most consistent with QGSJET which predicts the lowest mean mass in the knee region.  Thus this conclusion does not support the ideas of Erlykin and Wolfendale \cite{Erlykin} who favour an oxygen dominated cosmic ray beam in the knee region with iron becoming important at energies 26/8 higher. In their model most of the local cosmic ray flux is from a nearby supernova explosion. However a pattern in which the mass becomes lighter in the knee region before becoming heavier again is similar to what is predicted by Swordy \cite{Swordy} in the model mentioned above

\section{Cosmic Rays in the Region of the Ankle and above}

The first important feature in this energy range comes from the existence of radiation fields filling the universe of which the 2.7~K cosmic microwave background (CMB) is the best known. All stable particles expected to constitute the cosmic rays coming from large distances (except neutrinos) inelastically interact with those background photons above some threshold energy depending on the wavelength of the radiation and the nature of the cosmic rays: incident photons producing $e^+e^-$ pairs on infra-red radiation in the TeV range, on the CMB at a few hundreds of TeV, on radio waves at a few tens of EeV; incident protons with photopion production or nuclei undergoing photodisintegration at a few tens of EeV (the so-called Greisen-Zatsepin-Kuzmin, or GZK, spectral cutoff \cite{Greisen}). The important consequence of this is that at energies between, say, 50 and several hundreds of EeV, photons, protons and nuclei have rather short attenuation lengths, of the order of a few tens of Mpc. To state it more explicitly, if we observe cosmic rays in the above energy range on earth (let us call them ultra-high-energy cosmic rays or UHECR), it is impossible for them to come from a source whose distance would exceed 100 Mpc (roughly the size of our local supercluster), unless exotic particles or interaction models are envisaged.

The second feature comes from the chemical composition studies done by the AGASA \cite{Yoshida} and the Fly's Eye \cite{Bird} experiments (between 0.1 and 10~EeV). The Fly's Eye analysis is based on the measurement of the depth where the shower maximum $X_{\text{max}}$ is reached. The primary cosmic ray identification by AGASA (or more generally by ground arrays) mainly uses the muon content of the shower at ground level. A conclusion that can be reached from both analyses that the primary composition shifts from dominantly heavy -~compatible with iron nuclei~- to dominantly light -~compatible with protons~-  (see \cite{Dawson} for a critical review), even though discrepancies remain about the rate at which the change occurs. 

If this interpretation is to be believed (i.e. that the highest energy cosmic rays are mainly protons) there comes to light a third important fact. Although our knowledge of the galactic and extragalactic magnetic fields is not totally sound, a generally accepted working hypothesis sets the field strengths in the $\mu$G range for the galactic disk (with an exponential decrease in the halo) and the nG range (with coherence lengths of about 1~Mpc) in intergalactic spaces \cite{Kronberg}. The trajectory of a singly charged ultra-relativistic particle through such fields and over distances limited by the GZK cutoff is then such that its direction as measured on earth would roughly point to its source. Typically the angular deviation of a 100 EeV proton coming from a source at 30 Mpc would be of 2 degrees. Above the GZK cutoff, proton astronomy becomes possible to some extent, since the number of remarkable astrophysical objects inside a box of a few degrees and within a distance of a few tens of Mpc is quite limited.

A succint statement on the status of the highest energy part of the cosmic ray spectrum would be that our understanding of the origin of these cosmic rays becomes worse and worse as we go higher in energy. Due to the exponentially decreasing incident cosmic ray flux, the data around the `ankle' and above (roughly energies larger than $10^{18}$~eV or 1 EeV) come from a small number of large-aperture ground based detectors. Two types of detection techniques are used in this energy range. The first consists of ground arrays such as Haverah Park (United Kingdom) \cite{Lawrence}, Yakutsk (Russia) \cite{Afanasiev} and AGASA (Japan) \cite{Yoshida}, to which one should add the smaller Volcano Ranch (USA) array \cite{volcano} which detected the first air shower event to reach the symbolic limit of 100~EeV. The second technique, namely that of the fluorescence telescopes, is used in the Fly's Eye \cite{Bird} and the more recent HiRes \cite{Sokolsky} detectors, both in the state of Utah (USA). 

\begin{figure}[!ht]
\begin{center}  
\epsfig{file=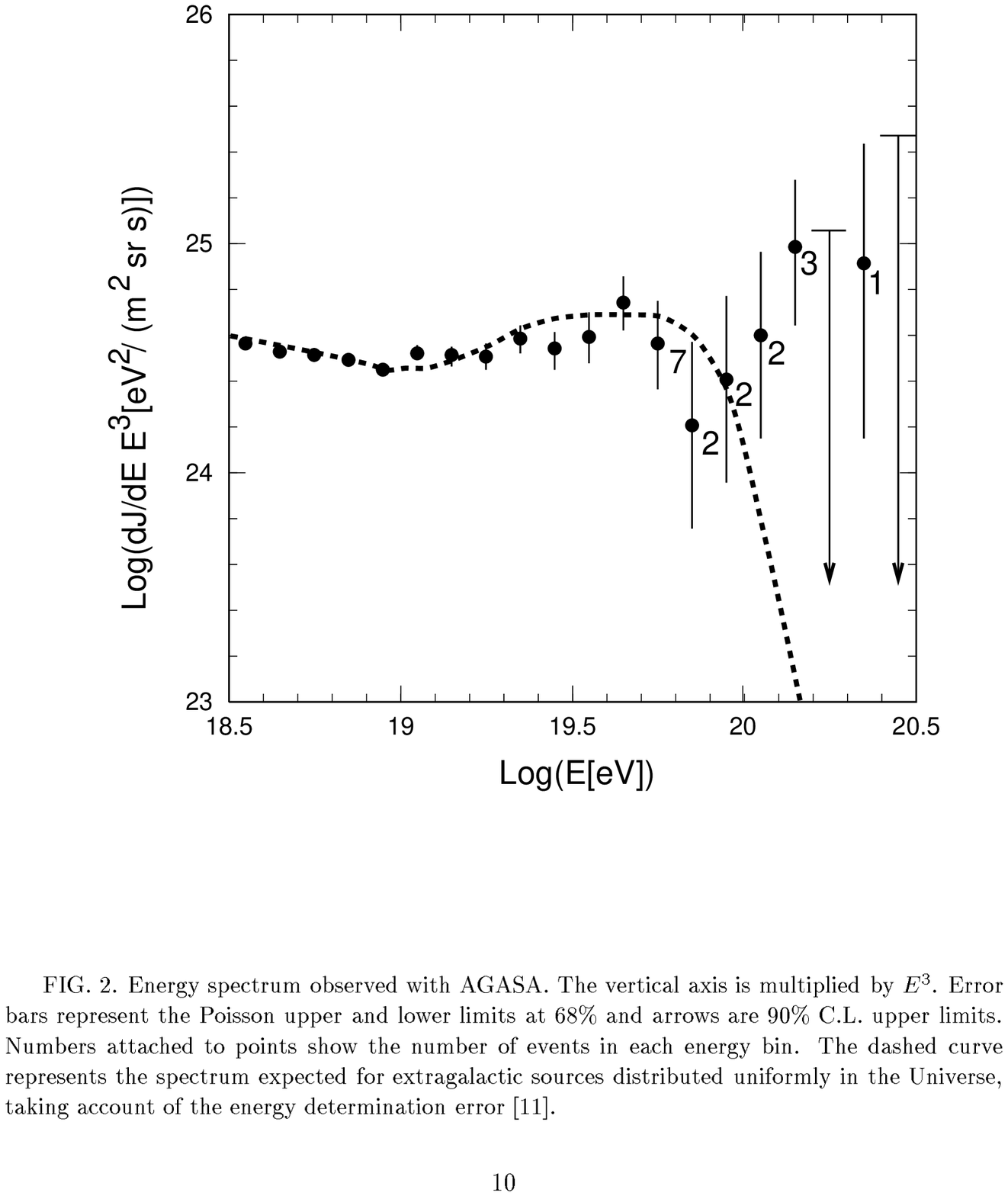,bbllx=117pt,bblly=228pt,bburx=512pt,%
bbury=602pt,width=8cm,clip=}
\end{center}  
\caption{\small Highest energy region of the cosmic ray spectrum as observed by the
AGASA detector \cite{Takeda1}. The figures near the data points indicate the
number of events in the corresponding energy bin. The arrows show 90\%
confidence level upper limits. The dashed line is the expected spectrum if the
sources were cosmologically distributed.\label{takeda}}
\end{figure}

The UHECR spectrum shape obtained by grouping all the available data points in a single figure is somewhat blurred because of the normalisation problems between different experiments. However the general trends are those clearly visible on figure \ref{takeda}. On this figure, where the AGASA data alone are plotted, 
the energy spectrum is multiplied by $E^3$ so
that the part below the EeV energies becomes flat. One can see the `ankle'
structure: a steepening around the EeV and then a 
region where the GZK cutoff is expected and shown by the dashed line which indicates the spectrum shape in a scenario where the sources are cosmologically distributed. The ultimate data points come from
very few events hence their large error bars. However, the apparent deviation of the events above 100 EeV from the model with cutoff is confirmed if one includes the 13 events detected by the other experiments (of which 7 events are claimed by the HiRes detector and not yet published).

The UHECR events constitute a puzzle, if not an enigma. Many astrophysical processes have been advocated in the past as being possible acceleration mechanisms capable of imparting a macroscopic energy to a microscopic particle (the most energetic cosmic ray ever detected has an energy of 50 joules!): active galactic nuclei, lobes of powerful radio-galaxies, young neutron stars, gamma ray bursts and so on (see e.g. \cite{Blanford} for a recent review). If all parameters related to the energy and chemical composition are taken into account -~acceleration, energy losses at source, propagation, detection~- one has to admit that none of the proposed scenarios seems fully convincing up to now. One has to keep in mind that because of the GZK cutoff, the putative sources need to be quite close and therefore should be visible by some counterpart in the direction of the highest energy incident cosmic rays. None of the analyses with the available data shows any strong correlation with known nearby point sources, small or large scale structures (see e.g. the AGASA analysis \cite{Takeda2}). However, this conclusion is challenged by a recent analysis \cite{Ahn} on the 13 highest energy cosmic rays. The authors propose a galactic wind model for the local magnetic fields and show that the observed cosmic rays can be back-traced to within 20$^{\circ}$ of the active galaxy M87 in the Virgo cluster (about 20 Mpc away). The only extra assumption they have to concede is that the two cosmic rays with the highest energies (200 and 320 EeV) are He nuclei instead of being protons. If confirmed by larger statistics from future data, this is a very exciting result and would open new windows not only in the field of astrophysics but also in particle physics: we would have a gigantic accelerator at hand reaching energies orders of magnitude higher than any conceivable man-made machine.

If it so happens that future studies exclude the ``conventional'' astrophysical mechanisms as being the source of the UHECR, one would then need to consider a second family of theories proposed as a possible explanation, namely the so-called ``top-down'' processes. Most of those study the possibility that the UHECR would be the decay products of some super-heavy X-particle whose mass is in the Grand-Unification range, i.e. around $10^{25}$ eV, produced during some phase-transition period of the early universe. The models differ mainly in their attempts to explain the density of the X-particles necessary to fit the observed UHECR flux and their survival from their production some $10^{-35}$~s after the Big-Bang until recent periods when their disintegration is supposed to occur. In a short review such as the present article, it is impossible to even list the various models which tackle with these thrilling issues (for a very complete review, see \cite{Sigl}) but we should mention that such models all have quite specific features and experimental signatures (spectrum shape and chemical composition) which make them not impossible to distinguish from the astrophysical (``bottom-up'') mechanisms, provided ongoing and future experiments in the field increase the available (scarce) UHECR data by some orders of magnitude.
\begin{figure}[!htb]
\begin{center}  
\epsfig{file=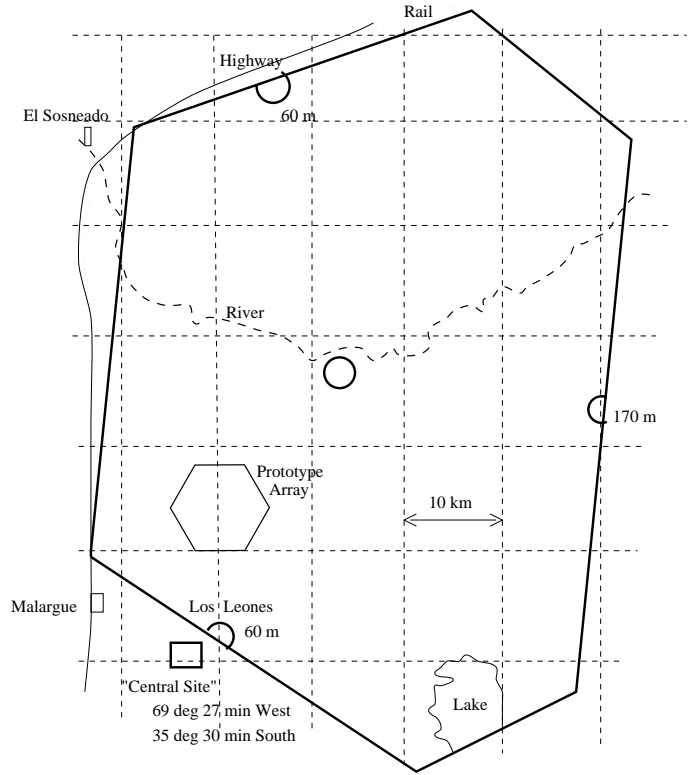,width=9cm}
\end{center}  
\caption{\small Layout of the southern Auger Observatory site (see text).\label{nihuil}}
\end{figure}

Actually statistics are the sinews of war in the search for answers to so many open questions in the field. This can be provided only by the next generation detectors which will provide the huge apertures needed. The next one will be the Auger Observatory \cite{auger} with its 14,000 km$^2$~sr aperture over two sites (one in each hemisphere). Once completed, it is expected to detect some 60 to 100 events per year above 100 EeV, and one hundred times more above 10 EeV. Figure \ref{nihuil} shows the layout of the southern observatory whose construction begins this year by the installation of a ``prototype'' array of 55~km$^2$ and one fluorescence telescope, near the small town of Malarg\"ue in the province of Mendoza, Argentina. \emph{In fine} the site will be equipped with 1600 detector stations (12 m$^3$ tanks filled with water detecting the Cherenkov light produced by the secondary shower particles) distributed on a grid with 1.5 km spacing. Four ``eyes'' (a total of 33 telescopes), three of them at the periphery and one at the centre, will view the 3000~km$^2$ of the site and detect the giant showers through the fluorescence they generate in the atmosphere during clear moonless nights. This setup will have the unique advantage of being a ``hybrid'' detector, combining the two aforementioned detection techniques. The surface array will detect air showers with energies in excess of 10 EeV with full efficiency and a 100\% duty cycle. A sub-sample of some 10\% of the total number of events will be simultaneously observed by the fluorescence telescopes, hence providing the possibility to cross-calibrate both methods and yield an unprecented quality for the identification of the primaries and the measurement of energy and direction.  

If in a few years it is shown that the spectral cutoff goes even beyond the reach of the Auger Observatory (e.g. $10^{21}$~eV or above) the next generation detectors will no doubt be air-borne as is envisaged with the OWL/Airwatch project \cite{owl}, fluorescence detectors installed on satellites.

\end{document}